\documentclass{article}
\usepackage{graphicx} 
\usepackage{orcidlink}
\usepackage{natbib}
\usepackage{authblk}  
\usepackage{url}
\title{Moving to ICSC: synergy between PNRR projects for more powerful Data Centers: a case study}

\author[1,2]{Fabio Ragosta}
\author[1]{Antonio Ferragamo}
\author[1,2]{Massimo Brescia}
\author[1]{Guido Russo}

\affil[1]{Dipartimento di Fisica “Ettore Pancini”, Università di Napoli Federico II, Via Cinthia 9, 80126 Naples, Italy}
\affil[2]{INAF - Osservatorio Astronomico di Capodimonte, Via Moiariello 16, I-80131 Naples, Italy}

\date{}
\begin{document}

\maketitle
\begin{abstract}
    The paper presents the new enhancement to the Data Center named DC1 at the University of Naples "Federico II". The ICSC funds at INFN have allowed to improve the 
power and cooling subsystems, while other funds from the PNRR (the STILES project) and funds directly from the MUR have allowed to enhance the computing, storage and network equipments. 
All these resources are in addition to the IBiSCo cluster and equipments described earlier in this book, but all together, thanks to a strong synergy between projects, have leaded 
to a very powerful Data Center for scientific applications. 
\end{abstract}

\section{Introduction}
The National Center for HPC, Big Data, and Quantum Computing, coordinated by the ICSC foundation\index{Italian National Centre for HPC, Big Data and Quantum Computing (ICSC)}, is designed 
to enhance research facilities and promote collaboration between the scientific and industrial sectors in Italy. The European Union program \emph{NextGenerationEU}\index{NextGenerationEU}, 
encourages the execution of cascade funding requests to support projects in Fundamental Research, Industrial Research, Experimental Development, and Feasibility Studies. 
The National Center for HPC, Big Data, and Quantum Computing is organized into 11 spokes\index{Spokes of ICSC}, the first of which (Spoke 0 or ``HUB'' is dedicated to infrastructure, 
while each of the remaining 10 focuses on a different theme.
Spoke 2 aims to provide a modular and domain-independent digital twin technology platform\index{Digital twin technology platform}. This platform should handle data heterogeneity, 
security, and privacy challenges utilizing open-source software and standards. It seeks to improve access to computing resources for present and future experiments, benefiting both 
the scientific community and the industrial sector. 

In the framework of this Spoke, at the University of Naples Federico II, we are developing a Data Center, which intends to transfer information and technology created via fundamental 
research with productive sectors, therefore, promoting the development of applications of national importance. Intersections with the rapidly increasing Space Economy sector will help 
to increase dissemination and technical transfer inside Italy.


The need for this type of infrastructure comes from the increasing complexity of scientific discoveries. These require ever larger collaborations presenting social and organisational 
challenges and yet at the same time offering the possibility of greater engagement with the public through outreach and industry. 
Large data centers have historically been the only places where data, to support scientific discoveries, have been handled. Hence a wider spread of data is now required. This demand stems 
from the request to have large data sets locally stored in the data center, provided by at least a mini- or mid-size Data Center. Moved by this demand, the Physics Department of the 
University of Naples ``Federico II'' 
in collaboration with ICSC-National Center of research in HPC, Big Data and Quantum Computing\index{Italian National Centre for HPC, Big Data and Quantum Computing (ICSC)}, creates a Data Center infrastructure. 

The Physics Department of the University of Naples Federico II is divided into multiple research divisions, suche as: astrophysics, particle and nuclear physics, theoretical physics, matter physics, 
biophisics; the majority of which have their own sets of scientific data that need to be stored and analysed. In this study we present the techniques used to implement a mid-size data center 
able to answer to this requests. A former data center was rebuilt in 2014 \cite[RECAS project]{russo2017recas}\index{ReCaS Project} after being started from scratch in 2007 (SCoPE project\index{SCoPE Project}). 

\begin{figure}
    \centering
    \includegraphics[width=0.8\linewidth]{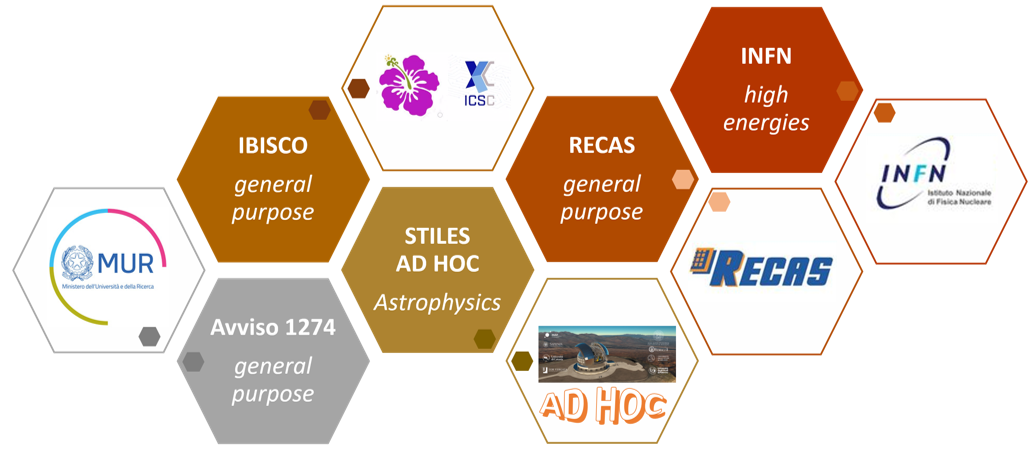}
    \caption{Scheme of the projects concerning the Data Center.
    }\label{C4-1:fig:dc1}
\end{figure}

The Data Center, indeed, has already been used to support research of scientific community proposing projects either of specific scope, or general purposes (see \autoref{C4-1:fig:dc1}), 
in a perspective of a transversal application of techniques for data analysis and storage.

\section{Implementation}
The infrastructure's architecture of the Data Center has been thought based on the data types the users have to work with. 
The DC1 structure contains, indeed, different clusters all dedicated to different researches for the scheme of the different clusters, among the different servers DC1 contains also the machines 
dedicated to IBiSCO\index{IBiSCo Project}, however the total computational power is at the disposal of all the users which are ranked through priority weights for the access to the servers capability. 

\subsection{Cluster STILES\index{STILES Project}}
Astrophysical scientists make up the first group of users; they gather and examine data from space telescopes (like Hubble and Webb), radio telescopes, and ground-based telescopes 
(such GranTeCan in the Canary Islands and ESO in Chile). These images need to be analyzed using AI technologies because they are usually rather large: between one and four million pixels each 
\cite[e.g.]{2010Ball}. 
In addition to image analysis tools that, with the right calibration, can be used to calculate physical attributes, AI will be able to identify different items in the files.

Material properties physicists represent another user group. They want to, for instance, create and test alternative jet fuels by providing a cutting-edge mathematical model that allows 
precise calculations of fuel consumption in jet engines \cite{Singh_2015}. 
This type of calculation needs the use of models with numerous variables, producing thousands of tests that must be saved and contrasted with one another.

The high energy physicists\index{High Energy Physics (HEP)} are a third category of consumers. The data analysis in a typical experiment (ATLAS\index{ATLAS LHC experiment} at CERN, 
BelleII\index{BELLE II experiment} at KEK) can be divided into two stages: first, selecting a sample of ``interesting'' events; and second, extracting the relevant quantities by 
analyzing the sample's overall properties and comparing them to the theory. The data collections span several dozen Terabytes or more \cite{Nairz_2014}.

\begin{figure}
    \centering
    \includegraphics[scale=0.35]{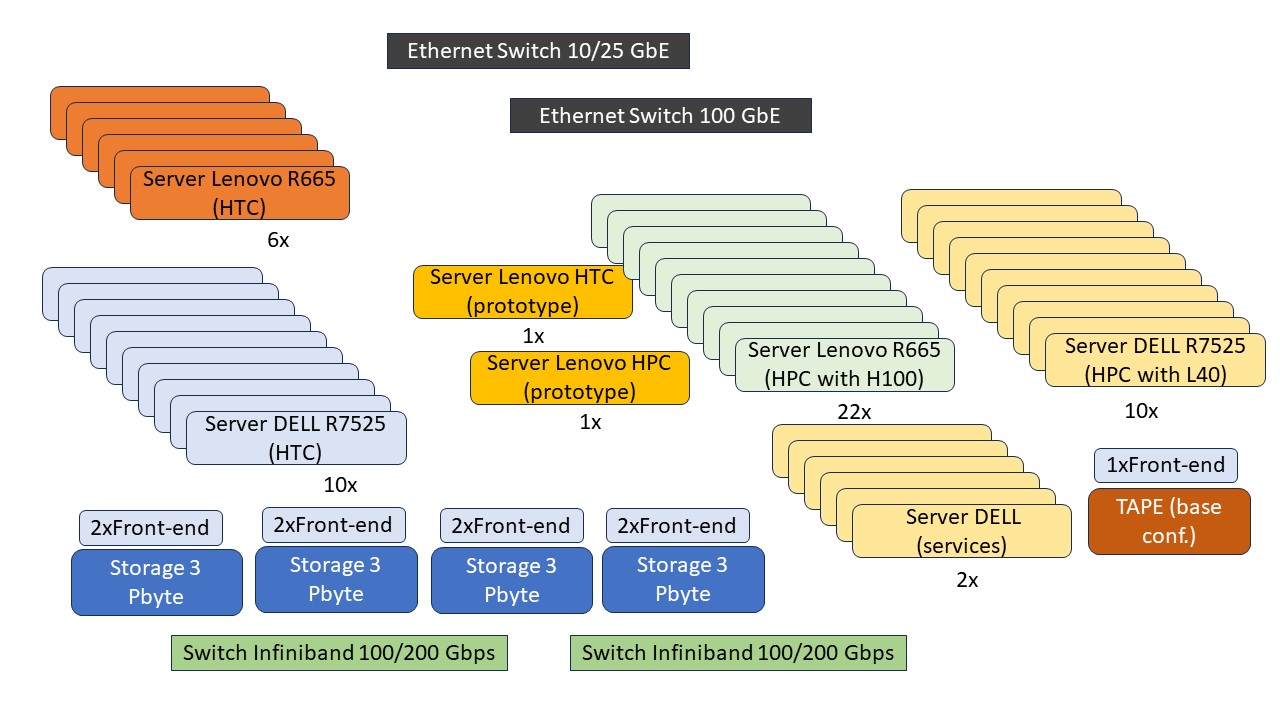}
    \caption{Configuration of STILES Cluster}\label{C4-1:fig:ClusterConfiguration}
\end{figure}

See Fig. \ref{C4-1:fig:ClusterConfiguration} for a representation of Configuration of STILES\index{STILES Project} Cluster.

\subsubsection{Storage}
\begin{figure}
    \centering
    \includegraphics[width=0.5\linewidth]{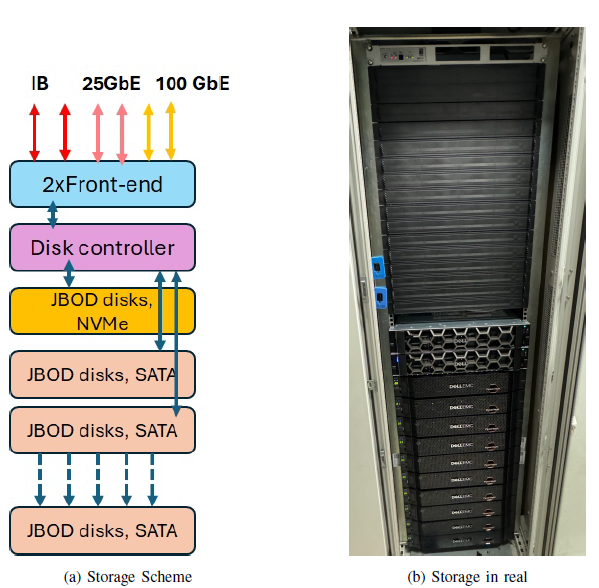}
    \caption{Architecture of the storage subsystem (replicated each 3 PBytes (raw)).}\label{C4-1:fig:scheme}
\end{figure}

The storage capacity needed is achieved with 500 SATA disks, 22 TB each, and 100 NVMe disks, 7.5 TB each (raw capacity). This bulk disk capacity is organized in groups of 12 enclosures, 
with several of these groups connected to a single Fibre Channel Arbitrated Loop (FC-AL) controller. The NVME disks act as a caching area to speed up access. Two server every 2.5 PBytes 
act as a front-end for the users, each server is connected to the controller via FC-AL. The server itself is accessible via multiple networking options: 2x25 GbE, 2x100 GbE, 1x200 
Infiniband \index{InfiniBand}. All the interfaces are duplicated towards different switches guaranteeing robustness and fail safe to the storage infrastructure. 
The file system is Lustre\index{Lustre file system} via Infiniband \index{InfiniBand}. Figure \autoref{C4-1:fig:scheme} shows a functional scheme of the system.

\subsubsection{Computation}
\begin{figure}
    \centering
    \includegraphics[width=0.5\linewidth,angle=-90]{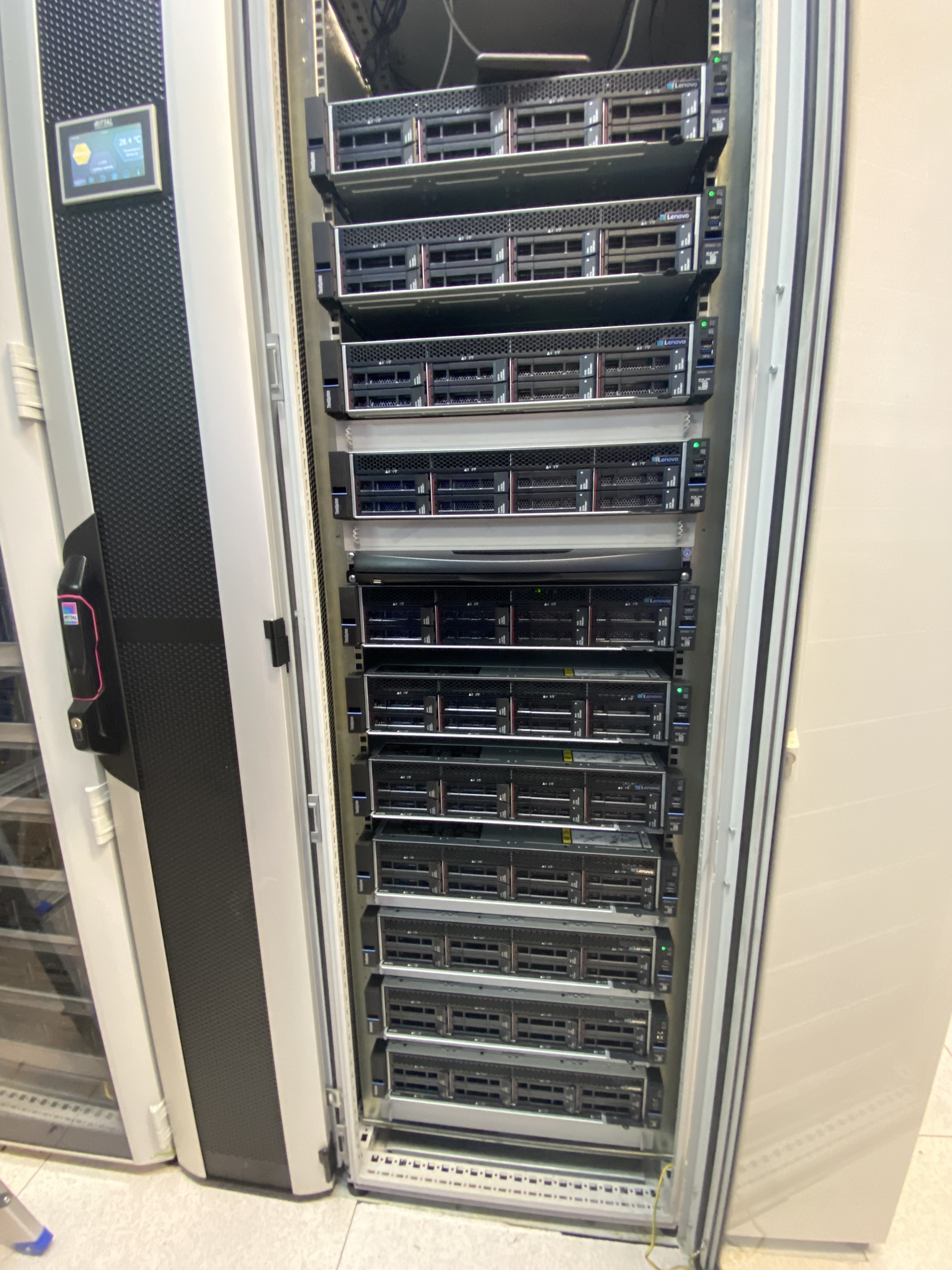}
    \caption{Some of the HPC Nodes in the DC1.}\label{C4-1:fig:nodes}
\end{figure}

For computational work, the Data Center has two types of nodes: HTC nodes\index{HTC Resources}, without GPU, and HPC nodes\index{HPC Resources}, with 2 GPU\index{GP-GPU Computing}.
In particolar, the hardware (part of which is shown in Fig.\ref{C4-1:fig:nodes} is the following:

\begin{itemize}
\item {HTC node {\em type A}: two 128 cores AMD, 3 Tbyte RAM, no GPU GPU, 2x25 GbE, 2x100 GbE, 1x200 Gb/s InfiniBand  \index{InfiniBand}}
\item {HTC node {\em type B}: two 32 cores AMD, 1 Tbyte RAM, no GPU, 2x25 GbE, 2x100 GbE, 1x200 Gb/s InfiniBand  \index{InfiniBand}}
\item {HPC node {\em type A}: two 84 cores AMD, 3 Tbyte RAM, 2xH100 GPU\index{GP-GPU Computing}, 2x25 GbE, 2x100 GbE, 1x200 Gb/s InfiniBand  \index{InfiniBand}} 
\item {HPC node {\em type B}: two 32 cores AMD, 1 Tbyte RAM, 2xL40 GPU\index{GP-GPU Computing}, 2x25 GbE, 2x100 GbE, 1x200 Gb/s InfiniBand \index{InfiniBand} }
\end{itemize}

There is also a group of HPC\index{HPC Resources} type C server, coming from the IBiSCo project\index{IBiSCo Project}, a third kind of computing node, that has been in 
operation since 2022. It is built on a server that has four NVIDIA V100 graphics cards\index{GP-GPU Computing} connected by NVLink\index{NVLink} and 0.77 Tbytes of RAM. Together, the 32 nodes 
which total 128 GPU\index{GP-GPU Computing} with EDR 100 Gb/S Infiniband \index{InfiniBand} comprise a cluster for general-purpose applications that is accessible to students as well.

\subsection{Cluster UNINA}
The Ministry of University and Research (MUR) has funded the enhancement of the scientific Data Center DC1 at UNINA through the {\it call for proposals} named {\em Avviso 1274}. 
Thanks to the synergy between the various projects, we have been able to buy exactly the same hardware components as the STILES project\index{STILES Project}. In particolar, 
we bought 12 HPC server, same configuration as STILES' HPC servers\index{HPC Resources}: 2x84 cores AMD processors, 3 Tbyte memory, 2xH100 GPU\index{GP-GPU Computing}, 
2x25 GbE, 2x100 GbE, 1x200 Gb/s InfiniBand \index{InfiniBand}. There are also 10 HTC servers\index{HTC Resources}: 2x84 cores AMD processors, 3 Tbyte memory, 2x25 GbE, 
2x100 GbE, 1x200 InfiniBand \index{InfiniBand}, but no GPU.

\subsection{The Network}
The Network subsystem in the DC1 is common to all the projects hosted in the Data Center, and each of themn has contribuetd to the necessary acquisitions.
There are three different Ethernet networks:

\begin{itemize}
\item The first network is at 10/25 GbE, made up with a stack of 6 Huawei ``CE-8861-4C-EI'' switches, each with foru boards and each with 96 SFP+ interfaces at 1/10/25 GbE and 8 QSFP28 interfaces at 100GbE. 
This is used for user access to all the machines. All ports are in fiber.
\item The second network is at 100 GbE, made up witch 2 Huawei ``CE-9860-4C-EI'' switches, joined not in stack but with the M-Lag protocol. This is used for all the machine-to-machine data exchanges, 
including MPI\index{Message Passing Interface (MPI)} and NFS. All ports are in fiber.
\item The third network is at 1 GbE, made up with 33 switches Zyxel GS-1900-48, with all ports in copper bur with fiber uplinks to a concentrator, an FS switch also at 1 GbE but with all ports in fiber. 
This network is used only for the management, and is not accessible by users.
\end{itemize}

\begin{figure}
    \centering
    \includegraphics[width=0.5\linewidth,angle=-90]{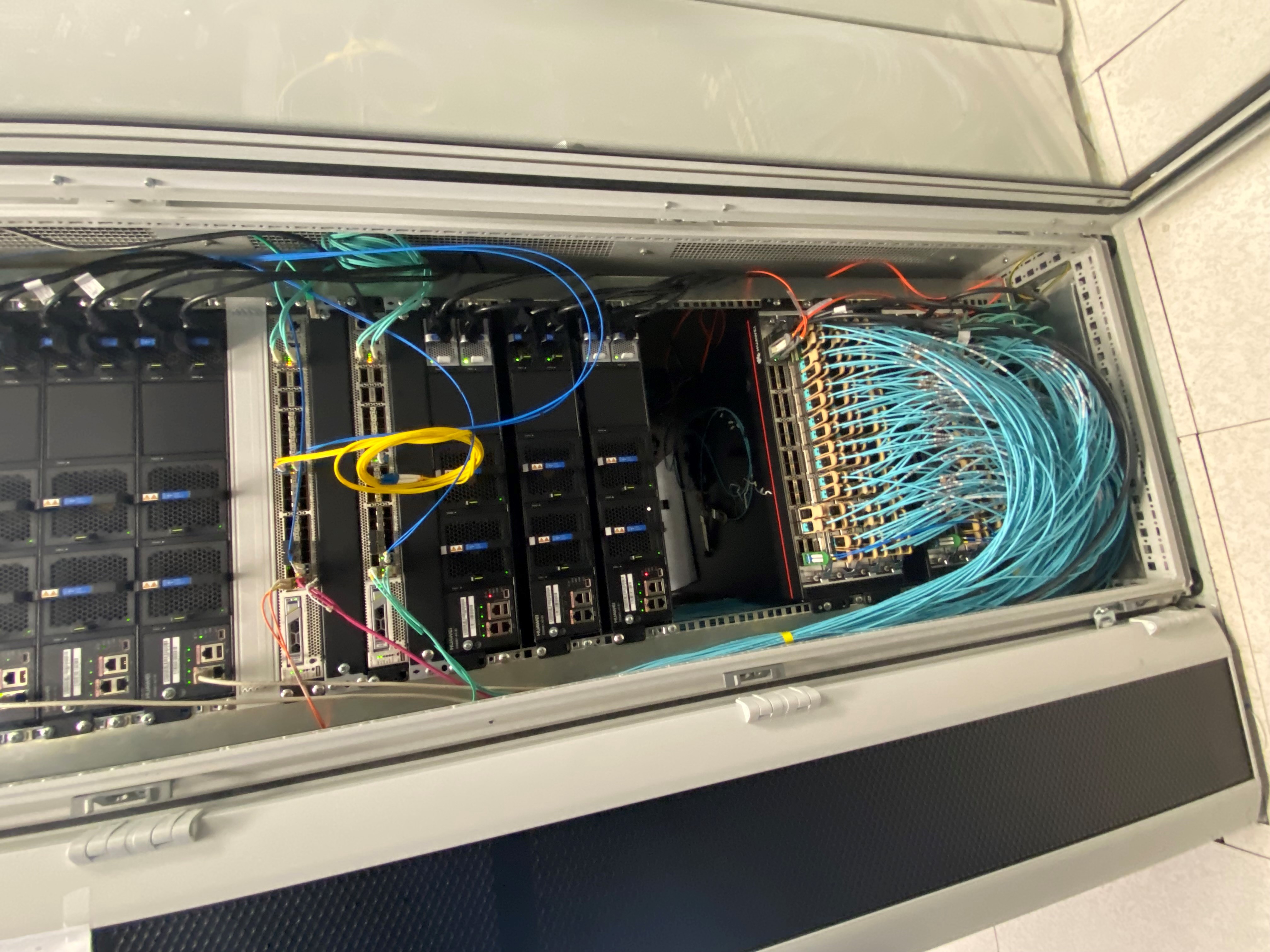}
    \caption{The rack (nr. 17) with all the switches in the DC1.}\label{fig:switches}
\end{figure}

The switches are all in rack 17, as shown in Fig.\ref{fig:switches}

\begin{figure}
    \centering
    \includegraphics[width=0.5\linewidth,angle=-90]{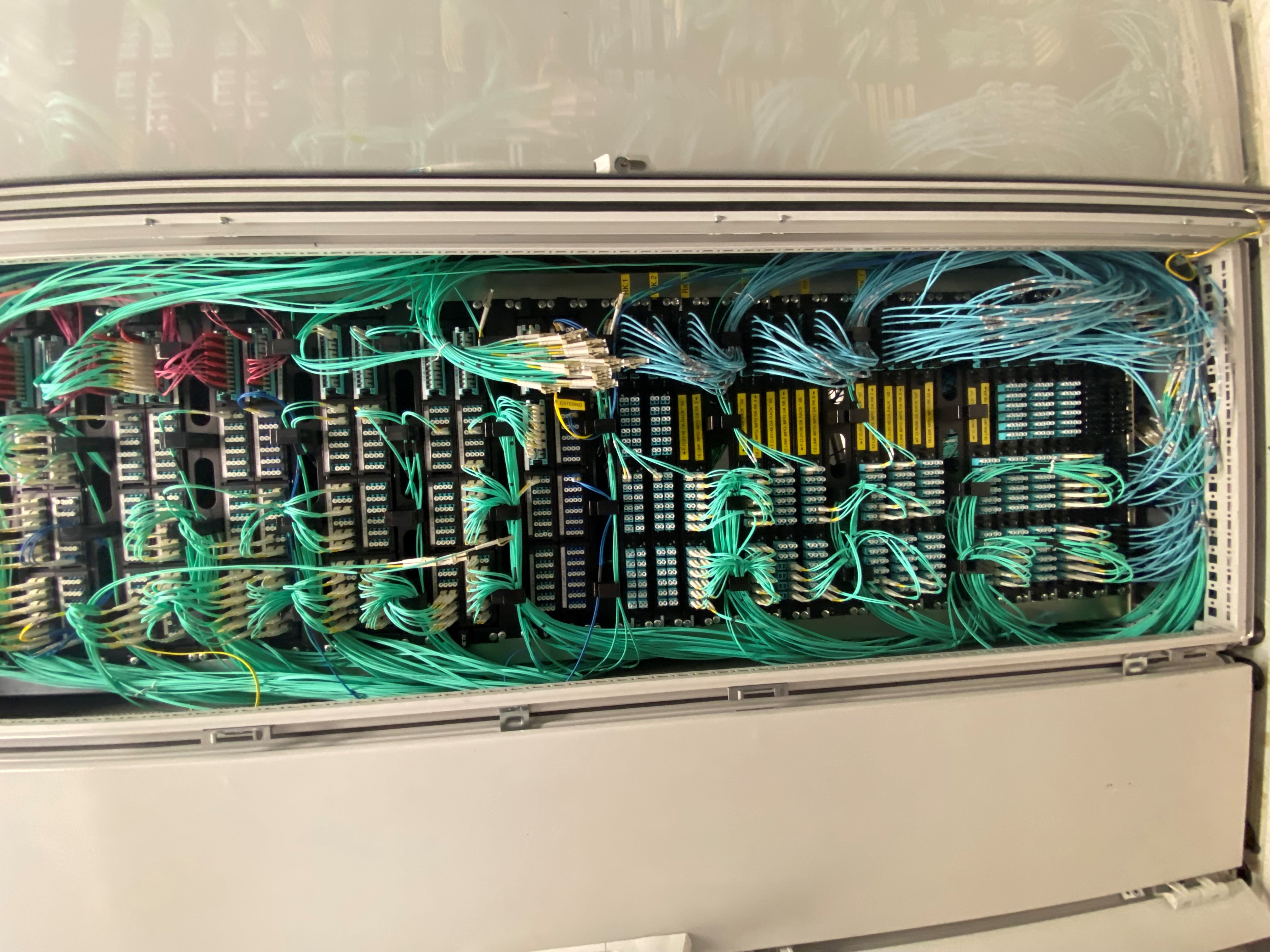}
    \caption{The rack (nr. 16) with all the fiber cabling in the DC1.}\label{fig:cabling}
\end{figure}

The entire fiber cabling system has been renewed during 2024, as shown in Fig.\ref{fig:cabling}

\section{Case study}
The Data Center as it is developed, it will support the AD HOC
\footnote{``Strengthening the Italian Leadership in ELT and SKA (STILES)'', proposal nr. IR0000034, admitted and eligible for funding from the funds referred to in the D.D. prot. no. 245 of August 10, 2022 
and D.D. 326 of August 30, 2022, funded under the program ``Next Generation EU'' of the European Union, ``Piano Nazionale di Ripresa e Resilienza'' (PNRR) of the Italian Ministry of University and Research (MUR), 
``Fund for the creation of an integrated system of research and innovation infrastructures'', Action 3.1.1 ``Creation of new IR or strengthening of existing IR involved in the Horizon Europe Scientific 
Excellence objectives and the establishment of networks''} 
(Astrophysical Data HPC Operating Center) infrastructure in the participation of the Italian scientific community to the V. Rubin Observatory\index{Rubin Observatory} activities and its 
related LSST (Legacy Survey of Space and Time) data survey \cite{2019LSST}. 

Rubin\index{Rubin Observatory} is an 8m class telescope characterized by a wide field of view and a high resolution, designed to carry out a survey of the so-called dynamic universe, 
mapping the entire sky visible in the southern hemisphere every few nights. This implies that approximately 20 TB of data will be collected each night. This poses extremely challenging aspects 
also from the technological point of view for the management and exploitation of the collected data. The camera will observe in six filters covering wavelengths from 320 to 1050 nm. 
The survey project foresees that about $90\%$ of the observing time is dedicated to uniformly observe about 800 times 18000 deg2 of the sky (adding up on all 6 bands), during 10 years 
and will produce a co-added map up to magnitude 27.5 in r band. In Figure 12 is possible to see a 10-years map tiled over the entire southern sky by Rubin telescope and its related LSST.

With the use of these data, a database (about 300 Petabytes in size) will be created that will support the majority of main research programs and contain roughly 37 billion observations of 
20 billion galaxies, 17 billion stars, and 6 million solar system objects. The primary pillar of the Italian participation is the in-kind contribution program, which consists of a number 
of scientific contributions from our community that have been formally agreed upon with the project Board. This program will return our scientists' data rights, enabling them to utilize 
LSST data products immediately rather than having to wait for the data to be made publicly available. Among the most intriguing pipeline contributions to the Rubin-LSST\index{Rubin Observatory} community 
is the investigation of crowded star fields (such as the Galaxy, Globular Clusters, and Magellanic Clouds) using specialized software. 

This software is extremely computing demanding and processing time consuming, requiring parallel computing paradigms as well as machine and deep learning methods aimed at characterizing 
stellar clusters and homogeneously deriving key parameters e.g., age, distance, reddening, metallicity, etc.) for globular and open clusters in the entire survey footprint.
Another crucial example of astrophysical use case is the support to the national research for two of the most important astrophysical projects of the next decade, ELT\index{ELT Project} and SKA\index{SKA Project}. 

\begin{figure}
    \centering
    \includegraphics[width=0.7\linewidth]{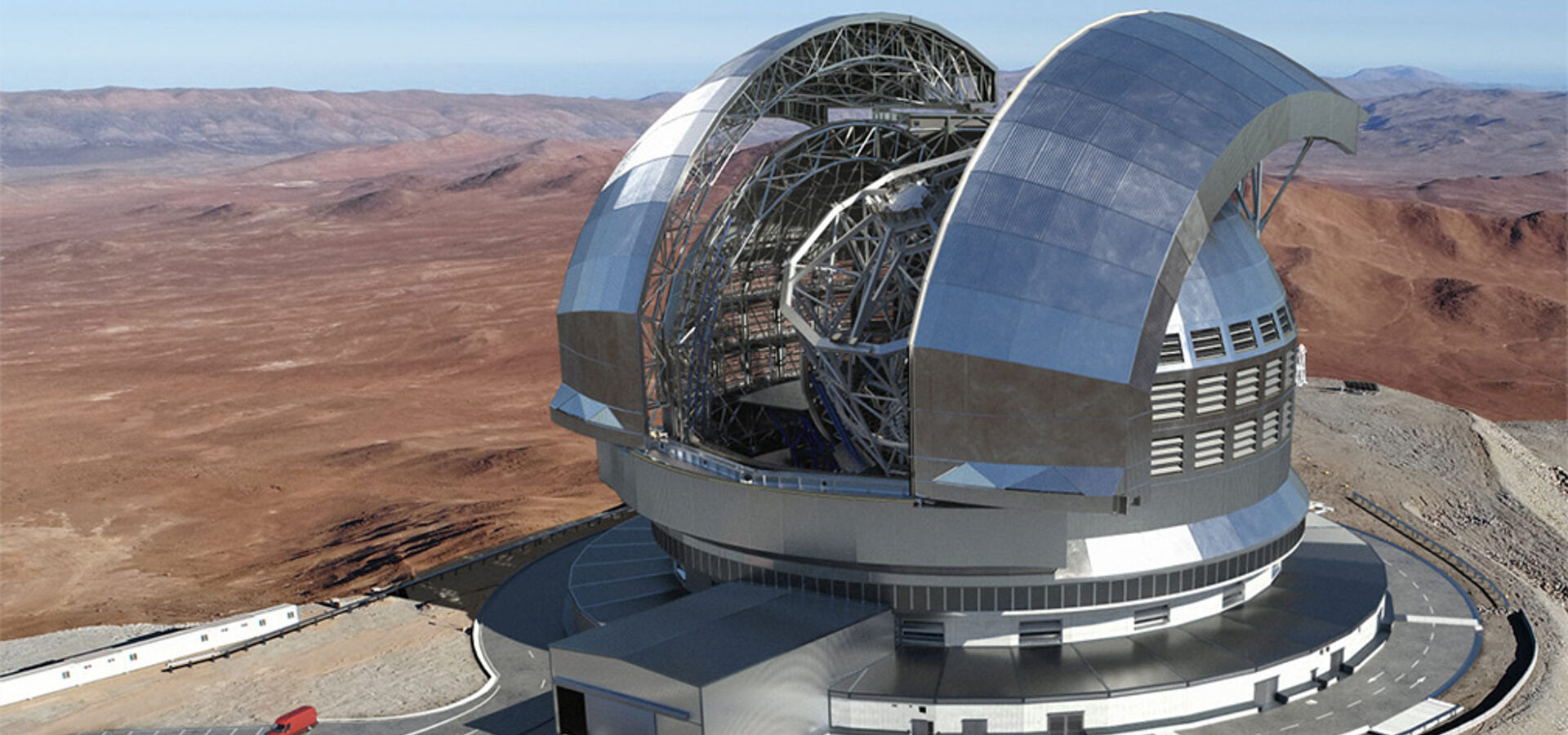}
    \caption{Artist's impression of the Extremely Large Telescope dome (ELT).}\label{fig:ELT}
\end{figure}

ELT\index{ELT Project} \cite{ELT} (Fig.\ref{fig:ELT}) is an optical/nearIR telescope with a 39m primary mirror, the largest of its kind ever built or planned. ELT is being built by ESO, 
and will be located atop Cerro Armazones, a 3000m peak in the Chilean desert. ELT is designed to exploit the full power of Adaptive Optics that removes atmospheric disturbances 
so as to reach the full resolution obtainable from the mirror and becoming able to produce data 5 times sharper and deeper than even JWST will do from space.

\begin{figure}
    \centering
    \includegraphics[width=0.7\linewidth]{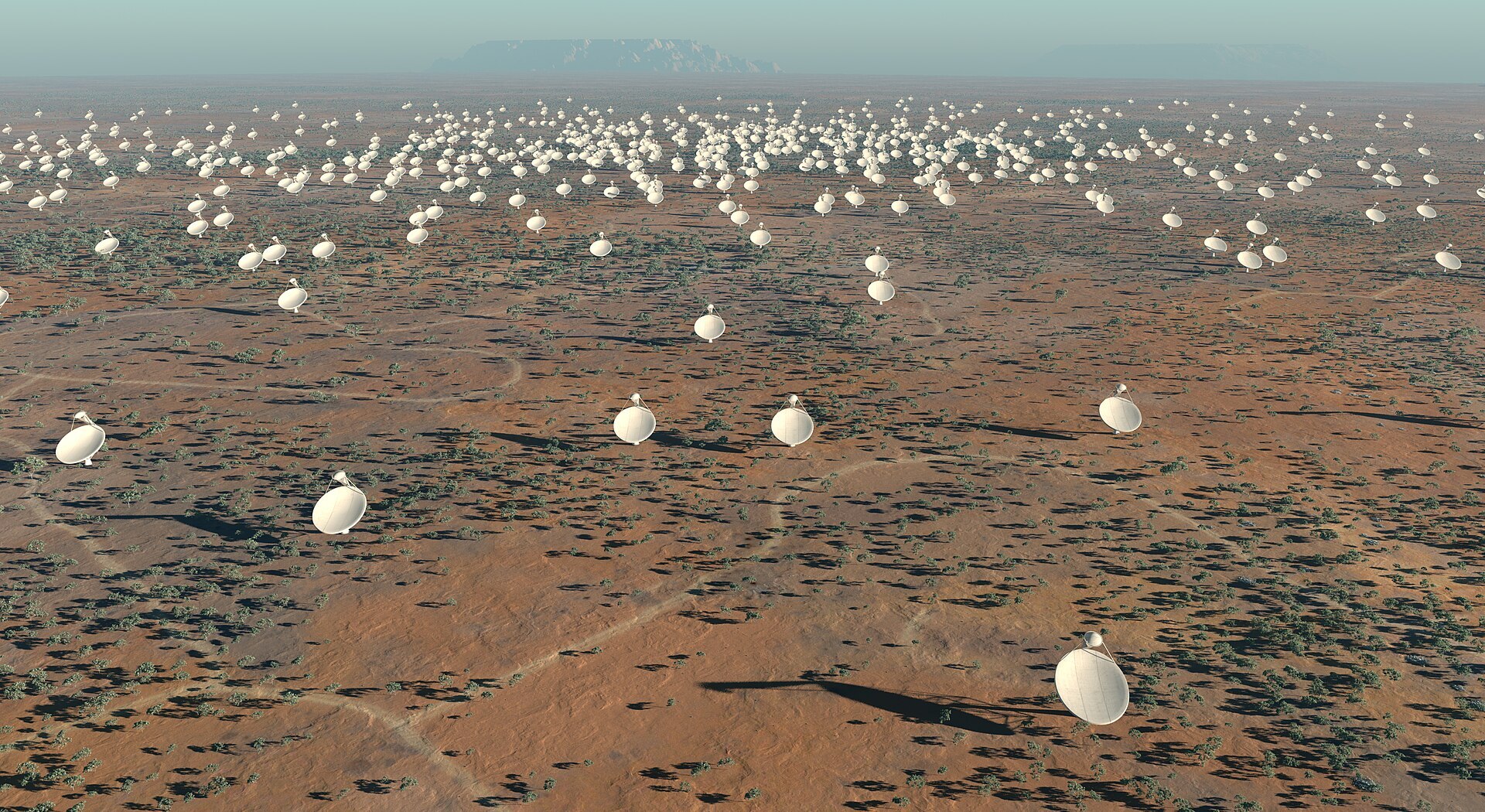}
    \caption{Artist's impression of the 5km diameter central core of Square Kilometre Array (SKA) antennas.}\label{fig:SKA}
\end{figure}

The SKA Observatory\index{SKA Project}\footnote{\url{www.skatelescope.org}} (Fig.\ref{fig:SKA}) will comprise two radio interferometers. The low frequency antenna array (SKA-Low) will be located in 
Western Australia and the mid-frequency antennas array (SKA-Mid) will be hosted in South Africa's Karoo region. 

SKA-Low will be made of 512 stations, each hosting 256 dual-polarized Log-Periodic antennas, distributed over an area of 65 km in diameter and operating over the 50-350 MHz range. 
SKA-Mid will comprise 197 dishes distributed over a region of 150 km in diameter, operating at frequencies from 350 MHz to 15 GHz, and will include the dishes of the precursor facility MeerKAT. 

The computing and storage support offered by our infrastructure will achieve a synergy between these two projects, since multi-frequency, multi-messenger approach is now recognized as a 
pillar of modern astronomy. Obvious examples of data products which will be supported by our computing infrastructure are the resource demanding multi-wavelength approach 

The ELT\index{ELT Project} will offer the exciting prospect of reconstructing the formation and evolution histories of a representative sample of galaxies in the nearby Universe by studying 
their resolved stellar populations (ESO - ELT Science).
Multi-wavelength observations are becoming mandatory to have a better interpretation of the models of these events. Hance, more observations are linked to very large volume of data to 
handle and to analyse: our goal here is to make available the computing facility and the analysis tools that are required to analyse data from ELT\index{ELT Project}, SKA\index{SKA Project} and other observatories, 
in synergy with the ESO archive and the SKA Regional Centres. 

The primary goal of the synergy between these projects will be to exploit Italian expertise for innovative data-mining techniques (generically referred to as Machine Learning). 
As already mentioned LSST survey project is a very important example for this case study. In this framework, we proposed an accurate PSF (Point Spread Function)\index{Point Spread Function (PSF)} reduction, 
based on a customised version of Allframe\index{Allframe Code} code \cite{stetson1994center}, of the most crowded areas in the sky (the Galactic Bulge and Plane, the 
MilkyWay globular clusters, the Magellanic Clouds, a few massive dwarf spheroidals). 

This kind of reduction is particularly demanding both on computing time and hardware resources. With the proposed new version of Allframe\index{Allframe Code}, a single master 
list of stars can be simultaneously measured on all the available images. This translates in a more accurate and fast photometry, since the positions of the centroids and the fluxes 
are cross-correlated among all the images. 

The typical approach for extracting flux from resolved stars is forced photomtry strategy, which is a technique used in astronomy to measure the brightness of an object at a specific location, 
using the object's known position to fit its flux even if it is not clearly visible in individual images. The proposed algorithm will be a game changer, since in the latter case the stars 
of the input list are measured individually on the target images. This different data mining strategy requires a HPC based\index{HPC Resources} computing framework, where, 
in particular, powerful GPUs\index{GP-GPU Computing} could be used to obtain the needed high level of parallelism to perform the simultaneous photometric reduction. 
The Allframe\index{Allframe Code} reduction needs a non-linear minimization of (N x M x S) matrix, where N is the number of images, M the number of the input stars and S the size 
of the involved CCD images, and it creates 2 working copies for each input image. Our typical run consists of the parallel reduction of the available DECam mosaics, each of them being 
made of 61 CCDs. The mosaics are reduced in about 15 hours, and this means that an individual 2k x 4k (S = 8 Mpixel) CCD is reduced in about 15 minutes. 

Expected number of stars per image is of the order M = 250k, while the number of images is N = 460. According to our experiments, a typical Allframe\index{Allframe Code} run with these numbers 
needs approximately 120 hours to be accomplished, and a single Allframe run needs approximately an allocated 1.5 Gb RAM. The disk space used by the process is dominated by the disk space needed 
for the images, that is 460 x 3 images x 31Mb = 42 Gb. 

The final catalogues are of the order of 7Mb for each CCD image, therefore 460 x 7Mb = 3 Gb for the whole set. With 189 4k × 4k CCDs in its camera, Rubin\index{Rubin Observatory}-LSST will 
yield around 850 visits for every single pointing at the end of the survey. Thus, we calculate that an Allframe\index{Allframe Code} run over 450 individual Rubin-LSST CCD images will 
need 3 GB of RAM, 85 GB of disk space for the input images and their working copies, and 6 GB of disk space for the permanent data products - or roughly 480 hours of single core computation time. 
The Rubin-LSST Camera's 189 CCDs must be multiplied by these figures. These numbers need to be taken into account as the largest possible ``final'' Allframe\index{Allframe Code} run size that can be parallelized.

\section{Conclusions}
We have described the realization of a computing center, mainly devoted to astrophysical applications, built up on a previous system, by replacing all energy-consuming devices with 
new powerful yet not-so-much expensive hardware and a simplified storage architecture with 10 Pbyte capacity. 

We have also presented some applications which would benefit by the computing power of the data center. 
The analysis of individual image from all the new generation ground based and space-borne telescopes will need the thousands of computing hours. Moreover, the consequent analysis of
all the data extracted by the observations and the storage of those data will be critical for the scientific community. 

A promising infrastructure for managing, analyzing, and storing large volumes of datasets is the ICSC\index{Italian National Centre for HPC, Big Data and Quantum Computing (ICSC)} Data Center. 
Our exercise taught us that the development of data centers and the deployment of HPC (High-Performance Computing) infrastructures\index{HPC Resources} do, in fact, indicate substantial 
improvements in data processing power. This is essential to ensure timely and effective analysis of the massive amount of data produced by the LSST and Rubin\index{Rubin Observatory} Observatory. 

Immediate access to LSST data products for Italian scientists promotes collaboration and expedites scientific discovery, which is a direct outcome. 

Deep learning, machine learning, and parallel computing paradigms are all being driven by the need for complex software and computational resources. This could result in the creation of 
novel technologies and algorithms that have uses outside of astronomy, helping other disciplines that depend on the processing of massive amounts of data. Effective utilization of 
HPC resources\index{HPC Resources}, such as strong GPUs\index{GP-GPU Computing} for parallel processing, maximizes data reduction and analysis efficiency. By doing this, processing 
times are shortened and management of huge datasets is made possible by ensuring that computational and storage resources are used as efficiently as possible.

\bibliography{C41}

\begin{thebibliography}{}
\expandafter\ifx\csname natexlab\endcsname\relax\def\natexlab#1{#1}\fi
\providecommand{\url}[1]{\href{#1}{#1}}
\providecommand{\dodoi}[1]{doi:~\href{http://doi.org/#1}{\nolinkurl{#1}}}
\providecommand{\doeprint}[1]{\href{http://ascl.net/#1}{\nolinkurl{http://ascl.net/#1}}}
\providecommand{\doarXiv}[1]{\href{https://arxiv.org/abs/#1}{\nolinkurl{https://arxiv.org/abs/#1}}}

\bibitem[{{Ball, Nicholas M. and Brunner, Robert J.}(2010)}]{2010Ball}
{Ball, Nicholas M. and Brunner, Robert J.} 2010, {International Journal of Modern Physics D}, 19, 1049, \dodoi{10.1142/S0218271810017160}

\bibitem[{{Ivezic, Zeljko and Kahn, Steven M. and Tyson J. Anthony and Abel, Bob and others}(2019)}]{2019LSST}
{Ivezic, Zeljko and Kahn, Steven M. and Tyson J. Anthony and Abel, Bob and others}. 2019, {Astrophysical Journal}, 873, 111, \dodoi{10.3847/1538-4357/ab042c}

\bibitem[{{M. Kissler-Patig and M. Lyubenova}(2009)}]{ELT}
{M. Kissler-Patig and M. Lyubenova}. 2009, {An Expanded View of the Universe, a high-level document outlining the E-ELT's science case}.
\newblock \url{https://www.eso.org/sci/facilities/eelt/science/doc/eelt_sciencecase.pdf}

\bibitem[{{Nairz, Armin and the ATLAS Collaboration}(2014)}]{Nairz_2014}
{Nairz, Armin and the ATLAS Collaboration}. 2014, Journal of Physics: Conference Series, 523, 012020, \dodoi{10.1088/1742-6596/523/1/012020}

\bibitem[{{Russo, G. and Barone, G.B. and Carlino, G. and Laccetti, G.}(2017)}]{russo2017recas}
{Russo, G. and Barone, G.B. and Carlino, G. and Laccetti, G.} 2017, in {High Performance Scientific Computing Using Distributed Infrastructures: Results and Scientific Applications Derived from the Italian PON ReCaS Project} ({World Scientific}), 57--71

\bibitem[{{Singh, Vedant and Sharma, Somesh Kumar}(2015)}]{Singh_2015}
{Singh, Vedant and Sharma, Somesh Kumar}. 2015, European Transport Research Review, 7, \dodoi{10.1088/1742-6596/523/1/012020}

\bibitem[{{Stetson, Peter B}(1994)}]{stetson1994center}
{Stetson, Peter B}. 1994, {Publications of the Astronomical Society of the Pacific}, 106, 250

\end{thebibliography}
\end{document}